\documentclass[aoas]{imsart}
\usepackage{amsmath}
\usepackage{graphicx}
\usepackage{enumerate}
\usepackage{url} 
\usepackage{mathtools}
\usepackage[dvipsnames]{xcolor}

\DeclarePairedDelimiter\floor{\lfloor}{\rfloor}
\usepackage{float}
\usepackage{enumitem}
\usepackage{amsfonts}
\usepackage[ruled,vlined]{algorithm2e}
\usepackage{algpseudocode}
\RequirePackage{amsthm,amsmath,amsfonts,amssymb}
\RequirePackage[authoryear]{natbib}
\RequirePackage[colorlinks,citecolor=blue,urlcolor=blue]{hyperref}
\RequirePackage{graphicx}

\startlocaldefs

\theoremstyle{remark}

\endlocaldefs

\begin{document}

\begin{frontmatter}
\title{Brain Waves Analysis Via a Non-Parametric Bayesian Mixture of Autoregressive Kernels}
\runtitle{}

\begin{aug}
\author[A]{\fnms{Guillermo} \snm{Granados-Garcia}\ead[label=e1]{guillermo.granadosgarcia@kaust.edu.sa}},
\author[B]{\fnms{Mark} \snm{Fiecas}\ead[label=e2,mark]{mfiecas@umn.edu}}
\author[C]{\fnms{Babak} \snm{Shahbaba}\ead[label=e3]{babaks@uci.edu}}
\author[C]{\fnms{Norbert} \snm{Fortin}\ead[label=e4]{norbert.fortin@uci.edu}}
\and
\author[A]{\fnms{Hernando} \snm{Ombao}\ead[label=e5]{hernando.ombao@kaust.edu.sa}}
\address[A]{
King Abdullah University of Science and Technology,
\printead{e1,e5}}

\address[B]{University of Minnesota,
\printead{e2}}

\address[C]{University of California Irvine,
\printead{e3,e4}}

\end{aug}

\begin{abstract}
The standard approach to analyzing brain electrical activity is to examine the spectral density function (SDF) and identify predefined frequency bands that 
have the most substantial relative 
contributions to the overall variance of the signal. However, a limitation of this approach is that the precise frequency and bandwidth of oscillations vary with cognitive demands. Thus they should not be arbitrarily defined \emph{a priori} in an experiment. In this paper, we develop a data-driven approach that identifies (i) the number of prominent peaks, (ii) the frequency peak locations, and (iii) their corresponding bandwidths (or spread of power 
around the peaks). We propose a Bayesian mixture auto-regressive decomposition (BMARD) method, which represents the standardized SDF as a Dirichlet process mixture based on a kernel derived from second-order auto-regressive processes which completely characterize the location (peak) and scale (bandwidth) parameters. A Metropolis-Hastings within Gibbs algorithm is developed for sampling from the posterior distribution of the mixture parameters. Simulation studies demonstrate the robustness and performance of the BMARD method. Finally, the proposed 
BMARD method was used to analyze local field potential (LFP) activity from the hippocampus of laboratory rats across different conditions in a non-spatial sequence memory experiment to identify the most prominent frequency bands and examine the link between specific patterns of activity and trial-specific cognitive demands.
\end{abstract}


\end{frontmatter}

\section{Introduction}
\label{sec:intro}

\begin{figure}[H]
\centering
    \includegraphics[width=14cm,height=8cm]{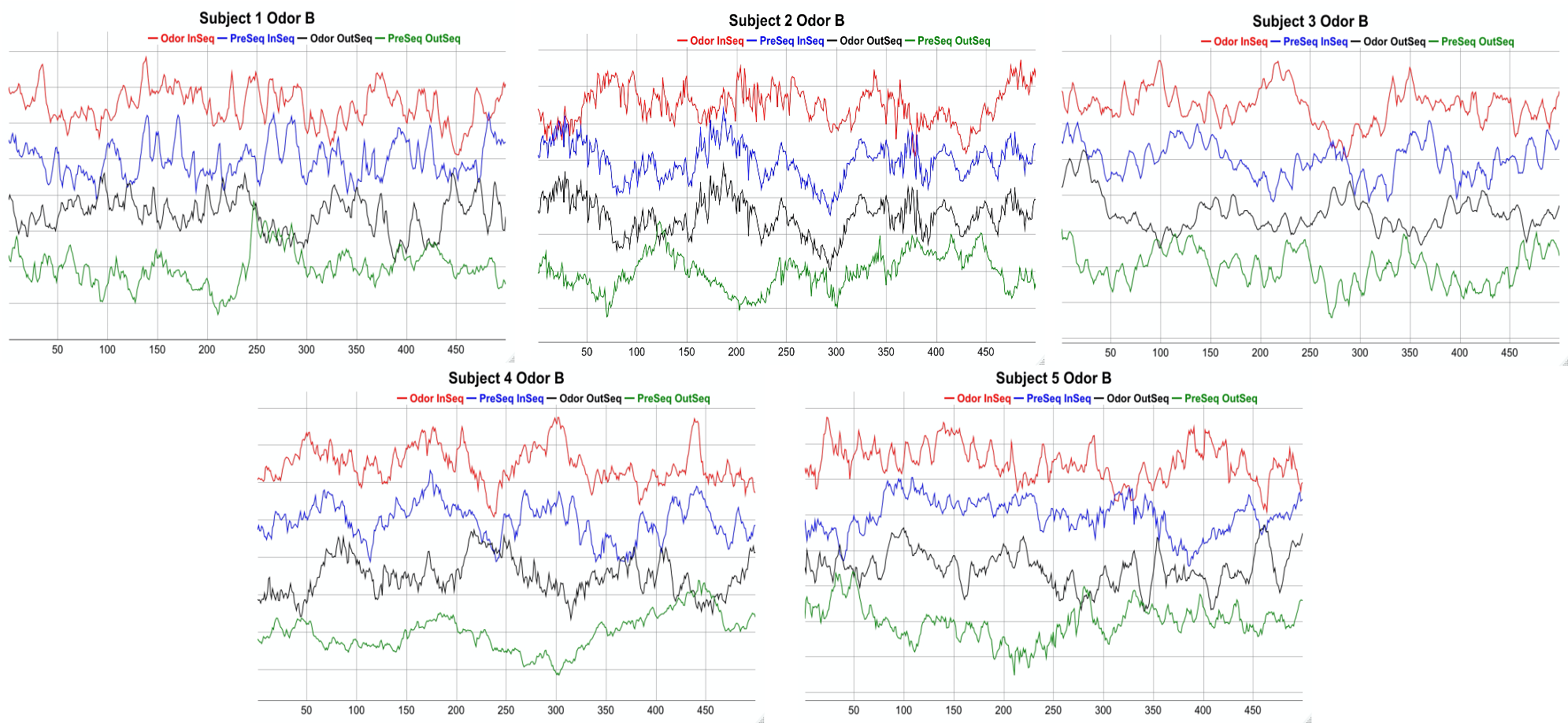} 
    \caption{Local field potential (LFP) signals recorded from the 5 rats during a odor sequence recognition experiment. We considered two phases of a trial: before the odor is presented ("PreSeq") vs. after the odor is delivered ("Odor"). There are two trial types: odor in a correct specified order ("InSeq") vs. odor presented in incorrect order (OutSeq).}
    \label{fig:LFP}
\end{figure}

Considerable research indicates that the hippocampus --- a brain region highly conserved across mammals --- plays a key role in our ability to remember the order in which daily life events occur \citep{eichenbaum2014time}. To identify the neuronal mechanism underlying this capacity, we have conducted an experiment in which neural activities are recorded in the hippocampus of rats as they perform a complex nonspatial sequence memory task.
Visualization of the local field potential (LFP) activity in Figure~\ref{fig:LFP} reveals a highly dynamic pattern of hippocampal oscillations during task performance, 
reflecting the distinct cognitive demands at different moments in time. Notably, the specific frequencies and bandwidth of the observed oscillations do not map well with standard predefined frequency bands (delta, theta, alpha, 
beta and gamma bands). Our goal in this paper 
is to 
identify in a data-driven manner the frequency peaks and 
bandwidth, and link these with specific types of information processing.
The frequency peaks and bandwidth (i.e., the most prominent 
oscillatory component of the signal) are captured by the spectral density function (SDF). Thus the focus of the present work is to estimate the SDF from which the most pertinent frequency information can be extracted. The SDF is a good descriptor of any stochastic process because it quantifies the amount of variability in a signal (such as
the LFP) that is accounted for by the different frequency bands 
\citep{shumway2017time, Brawnemery, ombaohandbook2017, prado2010time}. There are two general classes of methods for estimating the SDF. One class is based on the time domain where a parametric model, typically autoregressive moving average (ARMA), is fitted to the data and an estimate of the SDF is obtained by plugging in the estimates of 
the ARMA parameters. Another class is nonparametric and is based 
on kernel-smoothing or wavelet-denoising of the observed periodograms. 
The main limitations of the standard estimation methods and the general approach 
where the frequency bands are {\it a priori} defined are: (1) the lack of direct connection 
between the SDF estimators parameters to the time domain properties of the signal; (2) imprecise  location of frequencies with spectral peaks; and (3) the need increase the number of basic components of a model
to satisfy some estimation criterion (e.g., squared estimation error) which leads to an unnecessarily  more complex representation (or a less parsimonious representation) of the SDF. 
 
To overcome these limitations, we develop a nonparametric method, the Bayesian mixture auto-regressive decomposition (BMARD), by setting a Dirichlet process prior on the standardized SDF leading to a characterization of the SDF as a weighted mixture of kernels. We propose a kernel which is derived from the standardized SDF of a second order autoregressive process, that is associated to a unique oscillatory pattern. The BMARD approach leads to a representation of the signal as a weighted mixture of latent second order autoregressive processes where the number of components in the mixture is an unknown parameter in the model. For the BMARD estimator it will be sufficient to have a small number of kernels in order to obtain a suitable fit. Thus, the BMARD  provides a simple representations of the final SDF estimators that leads to easier interpretations and a more straightforward framework for addressing scientific questions related to the frequency and time domain properties of the data and perform experimental conditions comparisons. Most importantly, BMARD is not constrained by the 
a priori frequency bands used in standard analyses. It fully relies on the data to identify the location of the spectral peaks and also the frequency bandwidths of these oscillations.  

The remainder of this paper is organized as follows. In Section 2, we describe our model on Bayesian non-parametric framework through a Dirichlet Process prior and propose a new kernel based on the standardized SDF of a second-order autoregressive (AR(2)) process. The robustness of the BMARD method to model misspecification is 
demonstrated through simulation studies in Section 3. 
We also conducted a realistic simulation 
setting where the observed simulated LFP is a mixture 
of AR(2) latent processes with peaks that were actually observed in the the LFP signals recorded during the experiment in our co-author's laboratory. In Section 4, we analyze the hippocampal LFP signals of 
5 laboratory rats using the proposed BMARD method and address 
substantive questions in a non-spatial working memory experiment. The MCMC algorithm for obtaining posterior samples of the BMARD estimator is described in the 
Appendix.

\section{Bayesian Mixture Auto-regressive Decomposition}

The standard analysis for electrophysiological 
signals examines spectral power at frequency bands 
that are predefined, namely, delta (0.5-4 Hertz), theta (4-8 Hertz), alpha (8-12 Hertz), beta (12-30 Hertz), gamma ($>$ 30 Hertz). The power at the delta band indicates 
the contribution of slow oscillations to the total variance of the signal; whereas the gamma power is the contribution of fast oscillations.
The current segmentation of the frequency range into the delta-to-gamma frequency bands is ad-hoc and is primarily driven by pragmatic considerations \citep{buzaki}. However, in current research, many neuroscientists now consider some of these bands to be {\it too wide} and thus they do not possess the required level of precision in order to identify differences between stimulus types and between patient or treatment groups (see \cite{klimesch_induced_1998,doppelmayr1998individual}). There is an increased demand for more precise analyses with finer subdivisions within the bands for example as "low"-alpha and "high"-alpha (see \cite{Allen1547}, \cite{gao2016evolutionary}). Moreover, these frequency bands are predefined (according to the species under study) regardless of the experimental conditions. The proposed BMARD 
method will produce spectral estimates with peaks 
and bandwidths that are determined by the data -- rather 
than arbitrarily defined by the standard bands.

\subsection{Overview of spectral analysis}
To develop the specific ideas of our proposed approach, we first give a brief overview. Consider a process $X_t$ that is weakly stationary within an epoch with zero mean and autocovariance sequence 
$\{\gamma(h) = E(X_{t+h} X_t), \ \ h=0, \pm 1, \ldots \}$ that is  absolutely summable, i.e., $\sum_h \vert \gamma(h) \vert < \infty$. 
The SDF (within this stationary epoch) is formally defined as $f(\omega) = \sum_h \gamma(h) \exp(-i 2 \pi \omega h)$ where $\omega \in (-0.5, 0.5)$. Consider now the observed signal within an epoch $\{X_t\}_{t=1}^{T}$, where $T$ is even and 
the sample mean $\overline{X} = 0$. A nonparameteric estimate of the SDF $f(\omega)$ derived from the observed signal is derived from the Fourier periodograms $I(\omega_k) = \frac{1}{T} \vert \sum_{t=1}^{T} X_t \exp(-i 2 \pi \omega_k t)\vert$ which is computed at the fundamental frequencies $\omega_k =\frac{k}{T}$ (where $k \in \{-(\frac{T}{2}-1) \ldots, \frac{T}{2}\}$. 
The periodogram is an asymptotically unbiased for the 
SDF. However, it is not a consistent estimator because its variance does not decay to 0 even when the length $T$ of the observed process increases.

One way to construct a consistent estimator for the SDF $f(\omega)$ is by smoothing (or denoising) the periodogram. Several nonparametric methods have been proposed. 
Bandwidth selection methods for kernel and spline smoothing have been developed for this approach (see for example, \cite{lee1997simple}, \cite{ombaoGCV2001}, and \cite{wahba1980automatic}). These nonparametric methods aim to find a spectral estimator that minimizes a well-defined global criterion such as complexity-penalized deviance or integrated mean squared error. In addition, \cite{krafty2013} proposed a Whittle likelihood based approach. In \cite{krafty2011}, 
a functional mixed models approach was developed 
to account for the variation of the spectra in 
the setting with multiple processes data. 

An alternative class of methods is developed 
under the Bayesian framework. In \cite{CADONNA2017189}, a Bayesian method is proposed for estimating the log-SDF
which is modeled as a mixture of Gaussian distributions with frequency-specific means and logistic weights. A related approach, using the ideas in 
Bayesian nonparametric methods, uses kernels that 
are based on the Bernstein polynomial (BP). This 
was first proposed in \cite{petrone1999bayesian} 
to estimate a probability density function.  
The ideas were transported to spectral density function estimation in \cite{Choudhuri} and \cite{hart2018multi}, where the estimator 
uses a Dirichlet process (DP) mixture model with 
BP kernels. This was extended to the multivariate processes 
in \cite{macaro2014spectral}
which gives a decomposition in the frequency domain
in terms of BP-DP approximation. Furthermore, \cite{edwards2019bayesian} generalize the BP kernels with a procedure using B-splines prior thereby reducing the $L_1$-error. A method with BP-DP prior for spectral estimation with a nonparametric correction to the Whittle likelihood 
was developed in \cite{kirch2019}.
We note that mixture models based on Bernstein 
polynomials have the advantage that they are able 
to provide consistency in the pseudo posterior 
estimates. However, a limitation of these 
methods is that they do not offer a  data-generating mechanism in the time domain and can either oversmooth the peaks of the spectral estimates or require a high number of polynomials to achieve certain levels of accuracy.

A recent advancement on spectral estimation 
is the evolutionary state-space model (E-SSM) 
proposed in \cite{gao2016evolutionary}. The E-SSM offers a representation of the process as a mixture of second-order auto-regressive processes. The AR(2) SDF, derived from 
the phase and magnitude of the non-real 
complex-valued roots of the AR(2) polynomial function, are assumed to be constant in time. A property of this method is that the discrepancy between the true SDF and the approximate SDF derived from the AR(2) mixture vanishes by increasing the number of components. However, in practice, both the number of components and their phases are fixed. 
In some practical scenarios, it might be necessary to add components in order to 
produce better estimates.
This is a consequence of the rigidity when 
phases of the complex-valued roots are constrained to be fixed. However, in practice, a better solution is to adaptively identify the
locations of the various peaks in the SDF, which will produce a more parsimonious model 
and a more precise identification of spectral
peaks. The proposed BMARD method accomplishes these tasks. 


Nonparametric methods are flexible but, as noted,  their main disadvantage is that they generally do not have a straightforward data-generating mechanism. The parametric approach, on the other hand, is generally efficient but may not be always directly justified from the underlying brain physiology and may 
suffer from model misspecification. Here, our 
proposal, BMARD, is a Bayesian nonparametric approach for spectral density estimation that combines the best of both time and frequency domains. 

The BMARD, decomposes the SDF of a signal as a mixture of spectra of the AR(2) processes. Our approach provides a framework that describes precisely each specific oscillatory content in the signal.
More precisely, we model the standardized SDF as a multimodal probability density function by a DP-mixture model with kernels derived from the standardized SDF 
of second auto-regressive processes.
The advantage of the proposed BMARD is that it data-adaptively provides an estimate of the number of latent processes each with a unique location and scale parameters matching a single peak of the target standardized SDF. Unlike standard methods for spectral estimation, the BMARD method can precisely estimate the frequencies that produce the highest peaks in the standardized SDF without the need to constrain them to pre-specified bands, and determines the width of each of these peaks
(through a bandwidth parameter). Thus, BMARD  provides the practitioner with a more informative estimator of the oscillatory activity of brain signals and provides a more direct mapping between physiological signals and animal behavior (or 
cognitive response to various stimuli).

\subsection{Dirichlet Process Mixture Model}
\label{sec:meth}

Let $\{X_t, t=1, \ldots, T\}$ be the observed 
time series from s zero-mean weakly stationary process with SDF $f(\omega)$.  \cite{whittle1957curve} proposed a quasi-likelihood of the joint distribution of the periodogram values at frequency $\omega_k$,  denoted $I(\omega_k)$, expressed as the log-likelihood
 \begin{equation} \label{eq1wittl}
\begin{split}
\ell(f|X_1, \dots, X_T )= \sum_{k=1}^{\floor*{(T-1)/2}}    -\log (f(\omega_k))  -I(\omega_k)/f(\omega_k).   
\end{split}
\end{equation}
where $\omega_k= 2 \pi k/T, $ $k\in\{1,\dots, (T/2 -1)\}$. This quasi log-likelihood is based on the property that for $T$ sufficiently large, then  $\{I(\omega_k)\}$ are approximately jointly distributed as uncorrelated exponential random variables with
$E(I(\omega_k)) \approx f(\omega_k)$ for 
$k=1, \ldots \frac{T}{2}-1$. 

The formulation of the model, based on the periodogram values as input data, assumes that $f \ \sim \ F$, where $F$ is a probability measure specified given a parameter vector $\theta$ with unknown random probability measure $G$ associated to a Dirichlet process prior as follows:
\begin{gather*}\label{eqdpmodel}
I(\omega_k) \ \sim \frac{1}{f(\omega_k)}\exp{(-I(\omega_k)/f(\omega_k))}, \\
f(\omega_k)|\theta \sim F(\theta),\\
\theta|G \sim G,\\
G \sim DP(G_0,\alpha).
\end{gather*}
The prior $DP(G_0,\alpha)$ refers to the Dirichlet process \citep{ferguson1973bayesian,antoniak74,sethuraman1994constructive} with parameter $\alpha$ and probability measure $G_0$. Then for any finite partition of measurable sets $(S_1,S_2,\ldots,S_k)$ the probabilities $(G(S_1),\ldots,G(S_k))$ have a Dirichlet prior with parameters $(\alpha G_0(S_1),\ldots,\alpha G_0(S_k))$. In our case, due to the symmetry of the SDF, we consider a partition over the interval $(0, \ 0.50)$. The parameter $\alpha$ is a scale parameter of the DP that gives an 
indication on the number of estimated components. Low values of $\alpha$ leads to a posterior distribution of $G$ that is dominated by a few components 
(see \cite{muller2015bayesian,shahbaba2009nonlinear}). The prior $G_0$ is called a base measure and is associated to the prior distributions of the components of $\theta$.
The following papers give a detailed description on the role of $\alpha$ 
in determining the estimated number of components: \cite{gelman2013bayesian},  \cite{congdon2007bayesian}, \cite{neal2000markov}, \cite{shahbaba2009nonlinear}. 
In the implementation, the Dirichlet process is truncated up to $C$ components with a prior distribution over the positive integers, also called the truncated Dirichlet process (TDP), introduced in \cite{ishwaran2000markov}.

Note that in the usual DP-based clustering model, individual observations are assigned to each cluster with some probability. In contrast, our goal here
is to estimate the standardized SDF $g(\omega) = \frac{f(\omega)}{\int f(\omega) d\omega}$. Since  $\int g(\omega) d
\omega = 1$, $g(\omega)$ captures the shape of the SDF $f(\omega)$ and gives the proportion of variance explained by each frequency component. The standardized SDF $g(\omega)$ will be estimated using the 
periodograms derived by first standardizing the process so that they have unit variance. As a remark, note that these periodograms are not being "clustered" -- rather, they are associated with a convex combination of kernels.

The next section constructs a kernel that is based on a parametric model, where the shape of the 
standardized SDFs is a single peak represented in terms of a 
location (frequency peak) and scale (frequency 
bandwidth) parameter in the frequency domain. The new kernel allows us to specify the standardized SDF distribution, $F$, in the DP mixture model.

\subsection{Autoregressive Kernel and the DP mixture}
In the proposed model, we construct a kernel, $g$, as the {\it standardized} spectral density function (SDF) of an AR(2) process, leading to a  representation of the observed process as a linear mixture of multiple uncorrelated latent stochastic AR(2) processes each with unique spectra. 

\begin{figure}[H]
\centering
    \includegraphics[width=8cm, height=6cm]{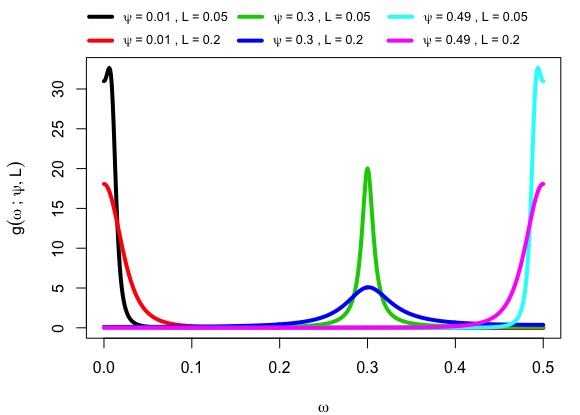}
    \caption{Some AR(2) kernels $g(\omega;\psi,L)$ with different values of the phase parameter $\psi=0.05,0.30, 0.49$ 
    as location (frequency peak) parameter and $L=0.05,0.20$ as scale (bandwidth) parameter.}
    \label{specexam}
\end{figure}

A weakly stationary process $Z_t$ is said to be  autoregressive of order 2, AR(2), if it 
has the representation
$Z_t-\phi_1Z_{t-1}-\phi_2Z_{t-2}=W_t$, where 
$W_t$ is a white noise process with variance $\sigma^2_W$ and the roots of the AR(2) polynomial function $\Phi(u) = 1 - \phi_1 u - \phi_2 u^2$ 
do not lie on the unit circle. 
Furthermore, when the roots of $\Phi(u)$ have magnitudes greater than 1 this AR(2) process is causal.  
When the roots, denoted as $u_1$ and $u_2$, are non-real complex-valued, then they are complex-conjugates of each other, that is, $u_1=u_2^*$ and $\vert u_j \vert >1$ ensures the  causality of the process. When $Z_t$ is causal 
with non-real complex-valued roots, then both roots have the polar representation 
\[
u_1 = M \exp(i 2 \pi \psi) \ \ {\mbox{and}} \ \ u_2 = M \exp(-i 2 \pi \psi)
\]
with magnitude $M >1$. The AR(2) polynomial function above is completely characterized either by the coefficients $(\phi_1, \phi_2)$ or by 
the roots $(u_1, u_2)$ or by the magnitude-phase of the roots through the following one-to-one relation between the roots and the coefficients in terms of the log-modulus $L=\log(M)>0$:
\begin{equation}
\phi_1=2\cos(2\pi \psi)\exp(-L), \qquad \phi_2=-\exp(-2L), \qquad \psi\in(-1/2,1/2), \qquad L>0.
\label{eq13}
\end{equation}
Due to symmetry of the SDF $f(\omega)$ at $0$, it is sufficient to specify the standardized SDF only at the frequency range $\omega \in (0, 0.50)$. In order to represent the SDF of $Z_t$ as a valid probability density function, it must integrate to 1 which is achieved by scaling by (or dividing by) $\int_{0}^{1/2}f(\omega) d\omega= Var(Z_t)/2=\frac{(1-\phi_2)\sigma_{W}^2}{2(1+\phi_2)((1-\phi_2)^2-\phi_1^2)}$. Then the {\it standardized} SDF is defined as 
\begin{equation}
g(\omega; \psi,L)= \frac{2(1-e^{-2L})((1+e^{-2L})^2-4\cos^2(2\pi\psi)e^{-2L} )}{(1+e^{-2L})|1- 2\cos(2\pi \psi)e^{-L}(e^{-i2\pi\omega}) +e^{-2L}(e^{-i4\pi\omega})|^2}
\label{eq19}
\end{equation}
where $\omega, \psi \in (0,1/2)$ and $L>0$. Here, $\psi$ is the location parameter of the kernel 
that attains a localized peak of the SDF of $Z_t$. The role of $L$ 
is to control the spread of the kernel as 
a scale parameter and hence it will be called 
the bandwidth parameter. Figure \ref{specexam} 
illustrates the roles of $\psi$ and $L$ in producing the different kernels.
A more formal justification for the selection of a second order autoregressive model is based on structural processes modeling ( see \cite{ozaki2012time}, \cite{Brockwell:1986:TST:17326}, \cite{shumway2017time}), where an AR$(p)$ process is represented in such a way that is equivalent to a consecutive input-output system of simpler processes. If the set of $p$ roots of its characteristic equation consists of $p_1$ real roots and $2p_2$ complex roots (here, $p = p_1 + 2p_2)$, then the characteristic equation is expressed as $(\Lambda -\lambda_1)(\Lambda -\lambda_1^*)...(\Lambda -\lambda_{p_2})(\Lambda -\lambda_{p_2}^*)...(\Lambda -\lambda_{p_2+1})...(\Lambda -\lambda_{p_1+p_2})$, where $^*$ denotes the complex conjugate. Then the AR$(p)$ can be represented through $p_1+p_2$ consecutive input-output system composed by $p_1$ AR(1) processes and $p_2$ AR(2) processes.
 Each pair of roots associated to an AR(2) process are constructed with following coefficients $\phi_1^{(i)}=\lambda_i+\bar{\lambda_i}$ and $\phi_2^{(i)}=-|\lambda|^2$ for $i \in 1,...,p_2$, while the real root associated to an AR(1) has coefficient $\phi_2^{(j)}=\lambda_j$ for $j \in p_2 +1,...,p_1+p_2$.  Our method propose to approximate the AR(1) components by AR(2) processes based on the the flexibility to fit simple SDF illustrated previously by estimating appropriate values for $\psi$ and $M$. Using this representation and the results in \cite{shumway2017time}, it follows that 
the SDF of a weakly stationary process can be approximated by a mixture of  AR(2) as latent factors, equivalent to a parallel structural model. 

In general for an arbitrary set of uncorrelated weakly stationary processes $\{Z_t^c, c=1, \ldots, C\}$ with corresponding identifiable autocorrelation function $\{\rho_{Z^c}(h), \ c=1, \ldots, C\}$ and identifiable SDF $\{g_c(\omega), \ c=1, \ldots, C\}$, their linear combination is still a weakly stationary process. Moreover, define $X_t = \sum_{c=1}^{C} a_c Z_t^c$ \ then we have the next relations: 
\begin{equation*}
 \sum_{c=1}^{C} a_c^2 g_c(\omega)= f_X(\omega) \ \ {\mbox{and}}  \ \ 
 \sum_{c=1}^{C} a_c^2 \rho_{Z^c}(h)= \rho_X(h) 
\end{equation*}
where $f_X(\omega)$ and $\rho_X(h)$ are the SDF and auto-correlation function of $X_t$ respectively. Conversely, a decomposition in the frequency domain may have this as 
one direct interpretation in the time 
domain. 

\subsection{Specification of the prior}
The definition of the parametric kernel completes the general specification of the DP mixture model state in equation \ref{eqdpmodel}. Now we discuss the prior distributions of the location and scale parameters of the kernel as well as the choice of a prior probability mass function for the number of components. 

In the practical implementation of the MCMC algorithm we avoid the potential label switching problem of the location parameter $\psi_c$, observed in mixture models, see \citep{jasra2005markov}, by defining a random partition over the interval $(0,0.50)$ as  $0<\epsilon_1<\dots <\epsilon_C$. We fix the first and last frequency block as $\epsilon_0=0$ and $\epsilon_C=0.50$. The partition protects the identifiability of the mixture components since we restrict each location parameter $\psi_c$ over the interval $\epsilon_{c-1}<\psi_c<\epsilon_c$ for $c=1,\dots,C$, thus constraining that there is only one component in each frequency block. 

At each iteration of the MCMC algorithm, the random cut off points $\epsilon_c$ are key to the procedure of generating the proper number of components. First, we randomly select a component label $c$ and propose to delete (death) or create (birth) with equal probability. When "birth" is chosen, we draw uniformly the candidate value for the partition cut off $\epsilon_*$ over the interval $(\epsilon_{c-1},\epsilon_{c})$ and propose randomly $\psi_*$ over the interval $\epsilon_{c-1}<\psi_c<\epsilon_*$. When  "death" is selected, then cut off $\epsilon_{c}$ is eliminated and a random $\psi_*$ is proposed.  

In the BMARD method, we select a prior for the bandwidth parameter $L_c$ that penalizes the full log-likelihood conditional on the DP mixture model parameters to capture sharp peaks. One example of the prior takes the form $L_c \ | \ C \sim L_c^\delta$; note here that $\delta =-2$ defines the Jeffrey's prior.
Utilizing this type of prior on the full conditional likelihood has some effect on the identification of the components as well as the smoothness of the components. In the M-H step, the resulting contribution of a new proposal $L^*$, relative a previous value $L$ on the log-likelihood, is $\delta \log(L/L*)$ which depends on the sign of $\delta$. If $\delta<0$ ($\delta>0$) sharper (broader) peaks update are penalized. However, if the update steps are small (i.e.,  $L^*=L+\epsilon)$ then this contribution will vanish due to the ratio $L/L*$ being close to 1. 

When the M-H step involves the creation of a new component, then the contribution of a new $L_{c+1}$ is $-\delta \log(L_{c+1})$; if the generated  $L_{c+1}$ is close to 0 then the contribution will be substantial. Then, for  $\delta<0$ the penalization is over a higher number of components; if $\delta<0$, it is over a smaller number of components. 
The prior for the number of components was chosen from the general form $\pi(c)=\exp(\lambda c^q)$ based on \cite{Choudhuri} which suggested $\lambda=-0.50, q=2.0$. This prior penalizes decreasing 
the number of components (in order to guarantee a more precise 
identification of peaks) since the contribution to the likelihood in the M-H step will be $\lambda(c^q-(c+1)^q )>0$. This prior will have more significant impact in cases when the observed signal is composed of low frequencies and when all the possible peaks 
are not well separated. Then the model could collapse to the simplest representation of one component and over-smooth the peaks. Allowing the number of components to increase as needed helps to detect and isolate peaks that are not easily distinguishable.

Given the prior definitions, the base measure $G_0$ of the DP Mixture model associated with the prior distribution over the parameter vector $\theta=(\psi, L, \epsilon)$, is as follows:
\begin{gather*}
\epsilon_c \ | \ \epsilon_{-c}, C \sim U(\epsilon_{c-1},\epsilon_{c+1}),\quad
\psi_c \ | \ \psi_{-c}, \bar{\epsilon}, C  \sim U(\epsilon_{c-1},\epsilon_{c}), \quad
L_c \ | \ C,b \sim L_c^\delta.
\end{gather*}
The posterior distribution of the DP mixing weights are estimated with the ``stick breaking" representation introduced by \cite{sethuraman1994constructive}. The posterior distribution of the parameter $\alpha$  is sampled based on two different priors: (i) $Gamma$ prior (as motivated in \cite{ishwaran2002approximate}) and updating the sample for $\alpha_i$ by a Gibbs sampling through the marginal $\alpha_{i+1}|V \sim \  {\mbox{gamma}}(M+a-1, b-\log(q_M))$, where $V$ is the vector of ``breaks" $V_j \sim \  {\mbox{Beta}}(1,\alpha)$, $q_M=V_M \prod_{i=1}^{M-1}(1-V_i)$ and M is the length of $V$, which is the level of truncation; (ii) log-normal prior and sampling from the distribution of $\alpha$ using a slice sampler \citep{robert2013monte} based on the posterior distribution given by \cite{escobar1995bayesian}.  

\section{Simulation Study}
\subsection{Setting and criteria}
Simulation studies were conducted to examine the relative 
strengths and weaknesses of the BMARD method  compared to three spectral estimation methods: (i) kernel regression smoothing using the Nadaraya-Watson estimator; (ii) a cubic smoothing spline approach and (iii.) the non-parametric Bayesian estimator developed by \cite{Choudhuri} based on Bernstein polynomials. Both (i) and (ii) are optimized with respect to leave one out cross validation (LOOCV) for the smoothness parameters.

\vspace{0.15in} 

\noindent {\bf The criteria.} In this study, three different parametric processes were used (see descriptions below). The methods were compared under the following criteria: (A.) The local integrated absolute error criterion (local IAE). Let $\omega_{max}$ be the
true value of the frequency at which the true SDF attains a peak and 
define $f(\omega_{max})$ to be the value of the SDF at the peak. Moreover, define a local interval around the peak to be $(\omega_{max} - \epsilon_1, \omega_{max} + \epsilon_2)$
where $\epsilon_1>0$ and $\epsilon_2>0$ satisfy $f(\omega_{max}-\epsilon_1)\approx 0.9 f(\omega_{max})$ and $f(\omega_{max}+\epsilon_2)\approx 0.9 f(\omega_{max})$. Then the local IAE of the estimator $\hat{f}$ around the frequency peak
$\omega_{max}$ is defined to be 
\begin{equation}
\mbox{Local IAE}_{\omega_{max}} = \int_{\omega_{max}-\epsilon_1}^{\omega_{max}+\epsilon_2} | \hat{f}(\omega)-f(\omega) | d\omega.
\label{localIAE}
\end{equation}
(B.) The maximal-phase disparity criterion. This criterion is inspired by \cite{dickinson2018peak}. Here, the focus is on identifying the frequency at which the SDF is maximized for a specific band -- as opposed to the local IAE criterion which focused on estimating the peak value of the SDF. This criterion was used in particular for the alpha frequency band motivated from cognitive studies 
where the alpha band is associated with learning. 
The absolute difference between $\omega_{max}$ and the location of the local maximizer of an estimator $\hat{f}$ within a frequency band $b$, i.e., $\hat{\omega}_{max}=\mbox{arg max}_{\omega \in b}( \hat{f}(\omega))$. The maximal-phase-disparity as between the true and estimated maximizer is defined to be 
$| \ \hat{\omega}_{max} \ - \ \omega_{max} \ |$.

Note that the integrated absolute error (IAE) computed through the whole frequency range as a global metric, i.e., it examines the performance across the entire range of frequencies. However, the local IAE measures the performance of the methods only in a local frequency range that contains the spectral peak. On the other hand the phase-disparity helps to evaluate the performance of the estimator to properly locate the peaks of the SDF.

\vspace{0.15in}
\noindent {\bf The simulation settings}. The first is a mixture of three AR(2) processes $Z_t^c$, \ $c=1,2,3$ with peak locations chosen similar to 
the ones observed in the LFPs in the data 
analysis. The frequency peaks were located at 8, 30, and 60 Hz (assuming that the sampling rate is $1000$ Hertz per second but we observe only half-second worth of data (i.e., $T=500$) which mimics the actual rat LFP data that will be analyzed later in this paper.
The peaks are associated to the following weights $p_1=0.1, p_2=0.6, p_3=0.3$. Moreover, $L_c=0.03,\ c={1,2,3,}$ which produce sharp peaks in the SDF. As noted, this construction simulates realistic brain signals since we observe a common component at 8 Hz, a second peak at 30 Hz, and the third peak represents an artifact at 60 Hz common to be found in brain signals measurements. Here, each pair of values $(\psi_c,L_c)$ defines a unique 
AR(2) process $Z_t^c$.

The goal of the next two settings is to test the robustness of BMARD with respect to a deliberately misspecified parametric model.
The second simulation setting is an AR(12) process that was studied in \cite{wahba1980automatic} to test the smoothing splines estimator of the standardized SDF. Moreover, the same setting defined in Equation~\ref{eq22} was also studied in \cite{Choudhuri} using Bernstein polynomials to estimate the AR(12) standardized SDF:
\begin{equation}
X_t=0.9X_{t-4} + 0.7X_{t-8} -0.63X_{t-12} + \epsilon_t
\label{eq22}
\end{equation} 
where $\epsilon_t$ is a white noise process. This process 
is useful to test the robustness of the BMARD method under model misspecification since 
the true process is not a mixture of $AR(2)$ processes. 
Here, we are examining how well the mixture can approximate 
a model with 3 main peaks at $\omega=0,250,500$ Hz and two smaller peaks at $\omega=150,350$ Hz. 
As a side remark, \cite{shumway2017time} explain that higher order AR models can approximate the SDF of any arbitrary stationary linear 
process. Here, the importance of estimating this type of model. The third setting is a MA(4) process generated as 
\begin{equation}
X_t=-.3\eta_{t-4}-.6\eta_{t-3}-.3\eta_{t-2}+.6\eta_{t-1}+\eta_t,
\label{eq23}
\end{equation}
where $\eta_t$ is a white noise process. Similar processes are discussed in \cite{wahba1980automatic}, \cite{lee1997simple}, \cite{ombaoGCV2001}, \cite{fan1996local}, and \cite{pawitan1994nonparametric}. The standardized SDF of this moving average process is a smooth curve centered at $\omega=250$ Hz with an extra bump around 500 Hz. This model would help to test our model under the scenario of fitting broad and smooth peaks, when the misspecification is not only in 
terms of the order but also in the structure of dependency since the MA processes have a zero correlation beyond the order (or when the absolute value of the lag exceeds the order).

To evaluate the performance of the BMARD 
method, 1000 processes were generated per 
setting each of $T=500$ time points. The data settings 
match the window size in the LFP data analysis. The spectral spline estimation was deployed using the package connection between R and C++: Rcpp \cite{rcpp1}, \cite{rcpp2}, \cite{rcpp3} in order to boost its efficiency, with the number of MCMC samples fixed to 100000 for six chains discarding 95000 as burn-in samples. 

The pointwise median of the sampled curves is computed considering 5000 after burn-in samples and all MCMC chains. The reported results are based on the implementation based on the gamma prior for parameter $\alpha$, the initial number of components was randomly selected in the set $\{1,\dots,20\}$ for each chain to start with different initial conditions aiming to convergence to the same posterior distribution. The level of truncation for the stick breaking representation was set random per each chain in the set $\{20,\dots,30\}$ as a conservative rule based on \cite{Choudhuri}.

\subsection{Results}

\begin{figure}[H]
\centering
    \includegraphics[width=14cm,height=10.5cm]{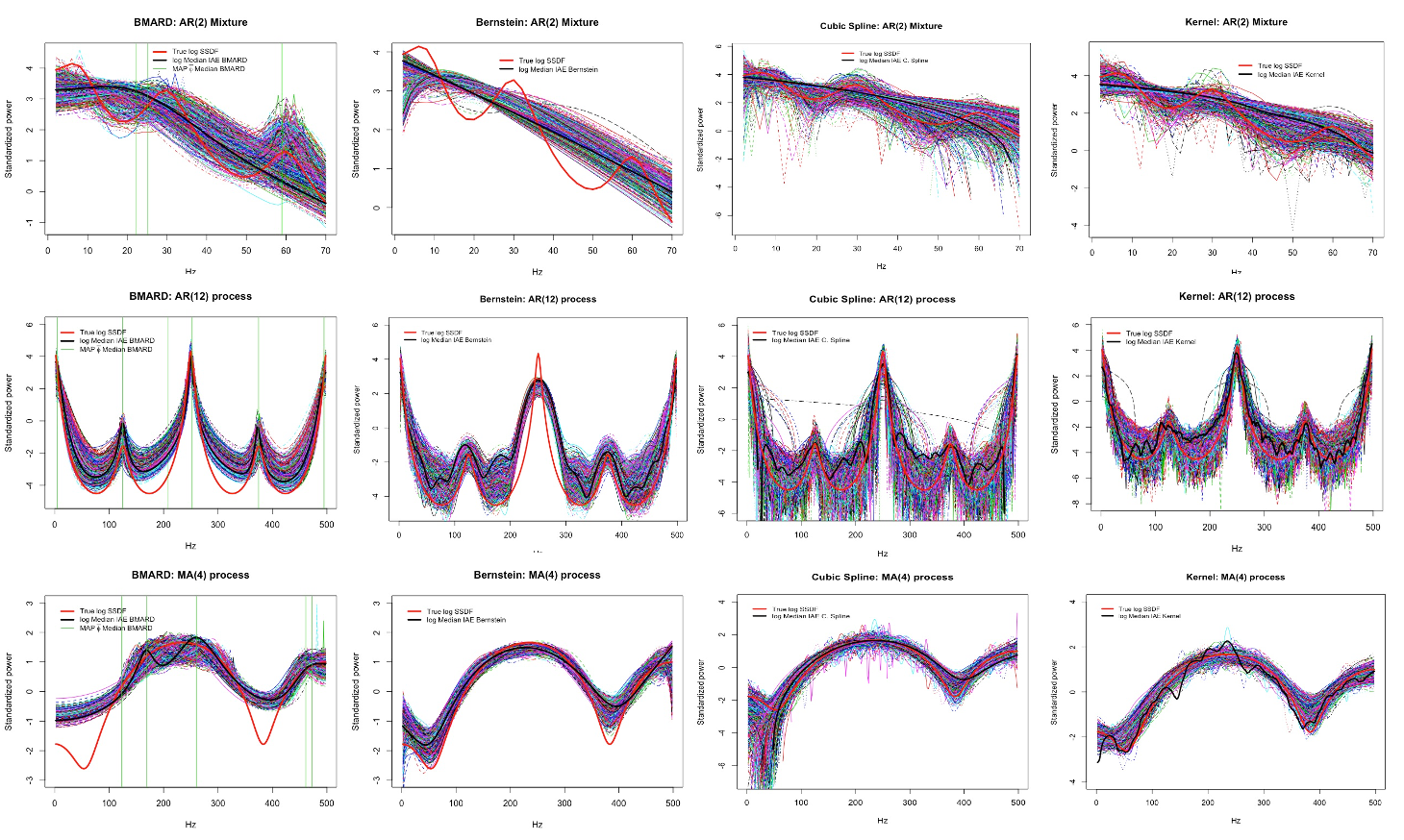}
    \caption{Estimated standardized SDF curves in log scale for all 1000 simulations highlighting in black the estimator with median IAE for the BMARD method included the MAP of the location parameters $\psi$ of the median IAE curve in order to visualize the components associated with the decomposition.}
    \label{logcurves}
\end{figure}

Figure \ref{logcurves} displays the logarithm of all the pointwise median curves per simulated processes for each of the methods. The first row corresponds to the AR(2) mixture setting. The results demonstrate that the BMARD method more accurately retrieves the peaks of the true standardized SDF compared to the other methods. It is also evident that the Bernstein polynomial method consistently smooths out the peaks. The spline and kernel estimators also identify the peaks in different simulations but, in general, produce higher variability across all frequencies. It capturing the main components but the estimated curves have more peaks than the true SDF.
 
The results from modeling the AR(12) process standardized SDF shows how the BMARD method outperforms the Bernstein polynomial method at retrieving the shapes of the peaks. The curves in log-scale indicate that when the SDF contains well-spaced and sharp peaks, the BMARD method produces better estimates. Regarding the algorithm sensitivity, we tested different values of $\delta$ and $\lambda$ for the bandwidth prior, and the level of truncation of the DP prior. The BMARD curves across chains generally consistently converged 
to similar curves and parameter settings.

In the MA(4) setting, the BMARD method was able to locate the peaks around the maximum of the true standardized SDF. However, it requires several components to approximate the smooth shape of the target standardized SDF due to the convexity of the autoregressive kernel - whereas the shape of the modes of MA(4) standardized SDF is concave. We point out this behavior of the BMARD method as a limitation if there is an interest in the shape around the peak (not only the actual location of the peak). The cubic spline estimator has a better performance followed by the Kernel smoother, showing as well higher variability on the curves observed in Figure \ref{logcurves}.

The local measures were compared mainly for the AR(2) mixture. The local IAE was similar for all methods, most likely because we sampled a window of only 500 points while the sampling rate is 1000 Hz, leading to compute the local IAE with only three frequencies within the band for each peak. The Bernstein method has a lower local error in the first two peaks, at 8 and 30 Hz, while the BMARD method and the classical smoothers behave similarly. For the last peak at 60 Hz, the BMARD method achieved a lower local error in several simulations. This pattern leads us to conclude that when the peaks occur in closely-placed frequencies, implying potentially indistinguishable periodogram peaks, the BMARD method tends to identify them as a single peak, which turns out in higher error. Of course 
this problem can be alleviated by increasing the 
number of observations but keeping the sampling 
rate fixed, i.e., by observing the data for a longer physical time. In contrast, when there is sufficient space between peaks (higher frequency resolution), the BMARD method can identify the peak activity even when the contribution to the total SDF is small. 

We use each of the single MCMC chain estimates of the parameters ($\psi, L, p$ ) to evaluate the average of the absolute for the maximal-phase disparity of the estimates compared to the true values. To this end, we consider only the chains that correctly estimate the true number of components to compute the average difference across the six chains over the 1000 simulations. Table \ref{tab:table1} reports the average absolute disparity and its standard deviation across the 1000 simulations for all the parameters of the individual components. The interval of the mean disparity plus two standard deviations contains 0 for the location parameters $\psi_c$ and the bandwidth parameters $L_c, c=1,2,3 $, which leads to conclude the estimation of both parameters is unbiased, for all components. However, the weights errors show the estimations differ from their true values, even when the shapes in the log scale of the mean curves in Figure~\ref{logcurves} demonstrate the estimated curves are close enough to the true SDF.

This behavior is due to having two parameters that contribute to the scale of the individual components of the mixture since the bandwidth $L_c$ at lower values narrows the Kernel implying a higher maximum. On the other hand, the weight $p_c$ directly shrinks or expands the scale of the $c-$th autoregressive kernel in the mixture. In further joint analysis of the parameters, we noted that the overall shape and contribution to the SDF estimation of the individual components is more sensitive to the bandwidth values.

To assess model fit, we investigated the convergence of the Whittle log-likelihood to ensure the stationarity of the MCMC for each of the generated datasets and each chain run. The evolution of the Whittle likelihood (not shown here) displayed an initial increase followed by a stationary behavior for all chains before reaching 50,000 iterations of the MCMC. Convergence of the Whittle log-likelihood 
to a stationary pattern was also observed for 
the Bernstein polynomial method. 
In summary, the simulations show that the 
BMARD method provides better estimates 
of the SDF when the true underlying process has auto-regressive (even higher order). 
However, it is less desirable than the 
nonparametric Bernstein polynomial method to fit the MA(4) SDF. Based on the AR(2) mixture simulations, 
the BMARD method was able to identify the SDF peaks more accurately compared with other methods as the BMARD estimated curves neither oversmooth nor overfit the periodogram. This higher accuracy of identifying peaks and bandwidth in the SDF is the central contribution of the BMARD, which  
directly addresses the current limitations of 
spectral analysis of electrophysiological signals in neuroscience.
\begin{table}[h!]
  \begin{center}
    \caption{Mean and standard deviation of the absolute difference with respect to the AR(2) mixture simulation parameters. Only for generated processes with at least one chain that correctly identified the true number of components. All components were generated with $L=.03$}
    \label{tab:table1}
    \begin{tabular}{c|c|c|c} 
       & $p_1=.1$   & $p_2=.6$ & $p_3=.3$\\
      \textbf{Mean Disparity} & $\psi_1=8$ Hz & $\psi_2=30$ Hz& $\psi_3=60$ Hz \\
      \hline
      $| \psi_c-\hat{\psi}_c |$ Hz& 7.65(5.17) & 11.66(10.32) & 7.96(8.76)\\
      $| L_c-\hat{L}_c |$ & 0.07(0.08) & 0.02(0.02) & 0.01(0.009)\\
      $| p_c-\hat{p}_c |$ & 0.6(0.22) & 0.37(0.15) & 0.23(0.05)\\
    \end{tabular}
  \end{center}
\end{table}
 \begin{table}[h!]
  \begin{center}
    \caption{Mean and standard deviation of the absolute difference with respect to the AR(12) peaks in the SDF. Only for generated processes with at least one chain that correctly identified the true number of components (5).}
    \label{tabAR12}
    \begin{tabular}{c|c|c|c|c|c}
      \textbf{ Mean Disparity} & $\psi_1=0$Hz & $\psi_2=150$Hz & $\psi_3=250$Hz & $\psi_4=350$Hz & $\psi_5=500$Hz \\  \hline
      $|\psi_c-\hat{\psi}_c|$ Hz& 4.77(2.19) & 32.67(21.36) & 3.94(14.38) & 31.09(19.58) & 3.68(2.07)
    \end{tabular}
  \end{center}
\end{table}

\section{Analysis of hippocampal LFPs from 5 rats} 

A major scientific goal in our collaborator’s research (Fortin laboratory, UC Irvine) is to understand how the hippocampus supports the ability of animals to remember the specific sequence in which events occurred which is a capacity critical to daily life function. While it is well-established that the hippocampus plays a key role in this capacity, we have little insight into how this is accomplished at the neural level. In particular, 
neuroscientists need to identify features in the 
observed signals (e.g., most prominent oscillatory 
patterns) that provide information about memory. 
To help achieve this goal, we apply the proposed BMARD method to the hippocampal LFP activity recorded from laboratory rats as they performed a nonspatial sequence memory task (similar to paradigms used in humans (see \cite{TAllen2014}). Indeed, the need for precise identification of oscillations in LFPs has been the primary motivation for developing the BMARD method. The second goal is to apply BMARD to detect stimuli-induced changes in brain signals through changes in the peak activity or shifts in the frequency content. 

\begin{figure}[H]
\centering
    \includegraphics[width=14cm,height=5.5cm]{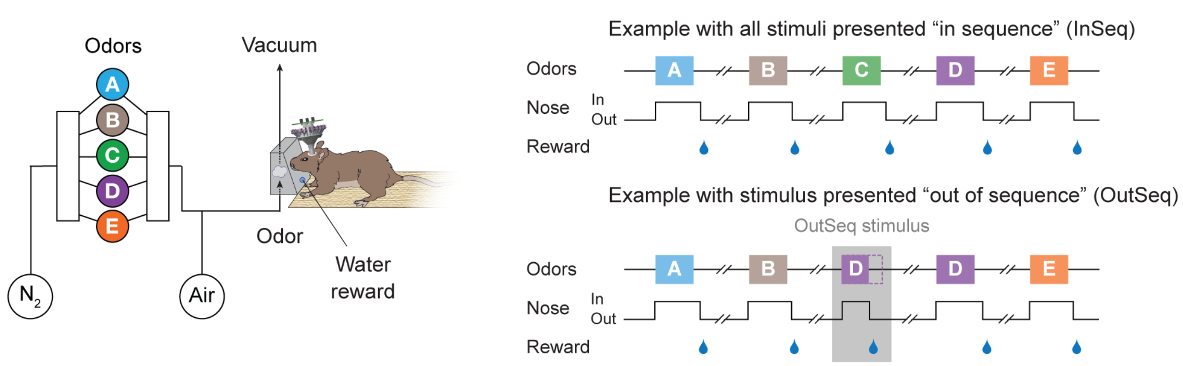}
    \caption{In this experiment, the animals receive multiple sequences of 5 odors (left; odors ABCDE). The animal is required to correctly identify whether the odor is presented "in sequence" (top right; by holding its nose in the port for $\sim$1.2 s, when an auditory signal is delivered) or "out of sequence" (bottom right; by withdrawing its nose before the signal) to receive a water reward.
    }
    \label{experiment}
\end{figure}

In the experiment performed in the Fortin lab, 5 rats (from now referred as subjects) were trained to recognize a sequence of five different odors (A = Lemon, B = Rum, C = Anise, D = Vanilla, E = Banana). A trial (i.e., a single odor presentation within the sequence) is labeled as "in sequence" (InSeq) if the odor is presented in the correct sequence position (e.g., ABCDE); otherwise, the trial is labeled as "out of sequence" (OutSeq; e.g., AB\underline{E}\ldots). Each subject is trained to identify InSeq trials by holding its nose in the port for 1.2s (when an auditory signal is delivered), and OutSeq trials by withdrawing its nose from the odor port before $1.2$-seconds, as illustrated in Figure \ref{experiment}. The LFPs were recorded from 20 electrodes (tetrodes) positioned in the pyramidal layer of the dorsal CA1 region of the hippocampus measured at a sampling rate of 1000 Hertz (1000 time points per second).

This data was first examined in \cite{Allen1547} where the authors analyzed the LFP activity using predefined frequency bands (4-12 Hz and 20-40 Hz). They found that, as animals ran toward the odor port, power was high in the 4-12 Hz band, particularly in the 7-10 Hz range. Upon odor delivery (when animals were immobile with their nose in the port), that oscillation seems to reduce in frequency (stronger in the 5-8 Hz range). Power in the 20-40 Hz range increased during odor presentations (particularly in the 19-35 Hz range), but was weak during the running period. Notably, 20-40 Hz power showed an association with session performance (higher in sessions with high performance). It also differed between InSeq and OutSeq trial types (higher on InSeq trials), although that analysis could not completely rule out the effects of uncontrolled differences in the animal’s behavior. 
Clearly, such dynamic and frequency-specific (narrow band) patterns analysis in the LFP is necessary but cannot be derived from standard analyses using broad, predefined frequency bands. 

To address this limitation, we use the BMARD posterior curves to conduct inference over all the frequencies and locate those with significant changes in power across experimental conditions. In the analysis, 
the first odor was omitted because, regardless of the sequence, odor A 
is always presented first (and hence always in correct order). In order to focus on the interaction between temporal context and sequence type, the trials in which a subject made the wrong response were excluded since different and more complex brain processes is expected to be present over wrong responses. The number of the trials during the experiment is displayed in table \ref{tabtrials}. 
 \begin{table}[h!]
  \begin{center}
    \caption{Total number of trials/epochs for each {\bf subject} (S1, S2, S3, S4, S5) for each {\bf odor type} (B-C-D-E) and {\bf stimulus type} (OutSeq vs Inseq).}
    \label{tabtrials}
    \begin{tabular}{c|c|c|c|c|c|c|c|c|c} \hline
      \textbf{Subject} & B InSeq & B OutSeq & C InSeq & C OutSeq &  D InSeq & D OutSeq & E InSeq & E OutSeq  \\  \hline
      S1   & 34 & 1 & 25 & 0 & 26 & 3 & 21 &2 \\ \hline
      S2   & 38 & 8 & 26 & 8 & 42 & 6 & 29 &7 \\ \hline
      S3   & 57 & 3 & 47 & 5 & 37 & 4 & 23 &4 \\ \hline
      S4   & 40 & 2 & 30 & 4 & 29 & 8 & 24 &2 \\ \hline
      S5   & 41 & 3 & 37 & 5 & 31 & 8 & 26 &5 \\ \hline
    \end{tabular}
  \end{center}
\end{table}

To identify the LFP dynamics associated with the processing of the odor stimuli we focused the analysis on a single electrode aligned at the same brain location for all subjects  for two time periods: a Pre-Odor baseline period (500 ms before odor presentation), and an odor period (focusing on the first 500 ms post-odor presentation, during which the animal's behavior is consistent across trial types). LFPs are generally non-stationary but it is reasonable to model each of the LFP records to be locally stationary and hence quasi-stationary  within the very brief intervals of 500 milliseconds. Separate analysis on a single electrode using BMARD consistently retrieved the peak activity observed in the observed periodograms. 
It is also of interest to consider the joint variability in oscillatory behavior across different tetrodes. Indeed there is a keen 
interest in the community to study potential lead-lag 
relationships between tetrodes and also various 
types of spectral dependence between pairs of tetrodes including coherence \cite{ombao2006coherence}, partial coherence \citep{park2014estimating,wang2016modeling}, and partial directed coherence \citep{baccala_partial_2001}. Future work will be on the  
generalization of BMARD for multivariate models. Under this framework 
we will develop methods for inference on 
cross-tetrodes connectivity - but this is outside of the scope of the current paper.
 
The BMARD was used to explore the posterior curves of each periodogram running eight chains of size 100000 considering a burn-in period of 90000 samples. We randomly set the initial number of components 
(of latent AR(2) processes) to be between 10 and 20. The sampler used for the posterior distribution of $\alpha$ is based on a gamma prior with initial parameters $a =0.10, b=0.10$ to set a less informative prior. We set the initial value of the chain $\alpha_0=1.00$. The level of truncation of the stick-breaking representation was selected randomly between 20 and 30. The final estimated curves were computed as the point-wise median of 10000 after burn-in posterior curves. With respect to the model fit assessment, we review 
the log-likelihood trace plots showing an increase and convergence to a stationary behavior for all the different trials and temporal contexts.


In our analysis, we used the LFPs at all trials and applied the two-stage approach for estimating the 
SDF separately for the Inseq and 
Outseq conditions. In the first stage, the SDF
was estimated separately for each trial; in the second stage, we combined information across 
trials within each of the Inseq and Outseq 
condition. There are many possible ways to obtain 
some "summary" across the SDFs including (a.) functional median curve \citep{ngo2015exploratory} (the 
option used here); (b.) point-wise median for each frequency; (c.) weighted average of all trial-specific SDF estimates where the weight is inversely proportional to the variance of the 
estimate (a spectral curve estimate derived from a 
very noisy trial should have low weight). Regardless, the summary tells us about the center of the distribution of true SDFs across all hypothetically infinitely many 
trials.

Some of the major questions posed in the Fortin laboratory that we attempt to answer using the BMARD approach are the following:  (a.) Are there differences in peak activity for pre-odor presentation vs post-odor presentation for the {\it Inseq} condition; (b.) Are there differences in peak activity for pre- vs. post-odor presentation for the {\it Outseq} condition; (c.) Is there potential interaction between condition (Inseq vs Outseq) and temporal context (pre-odor vs post-odor presentation), i.e., 
is the difference between pre vs post-odor 
the same across both the Inseq and Outseq 
conditions?

\begin{figure}[H]
\centering
    \includegraphics[width=13cm]{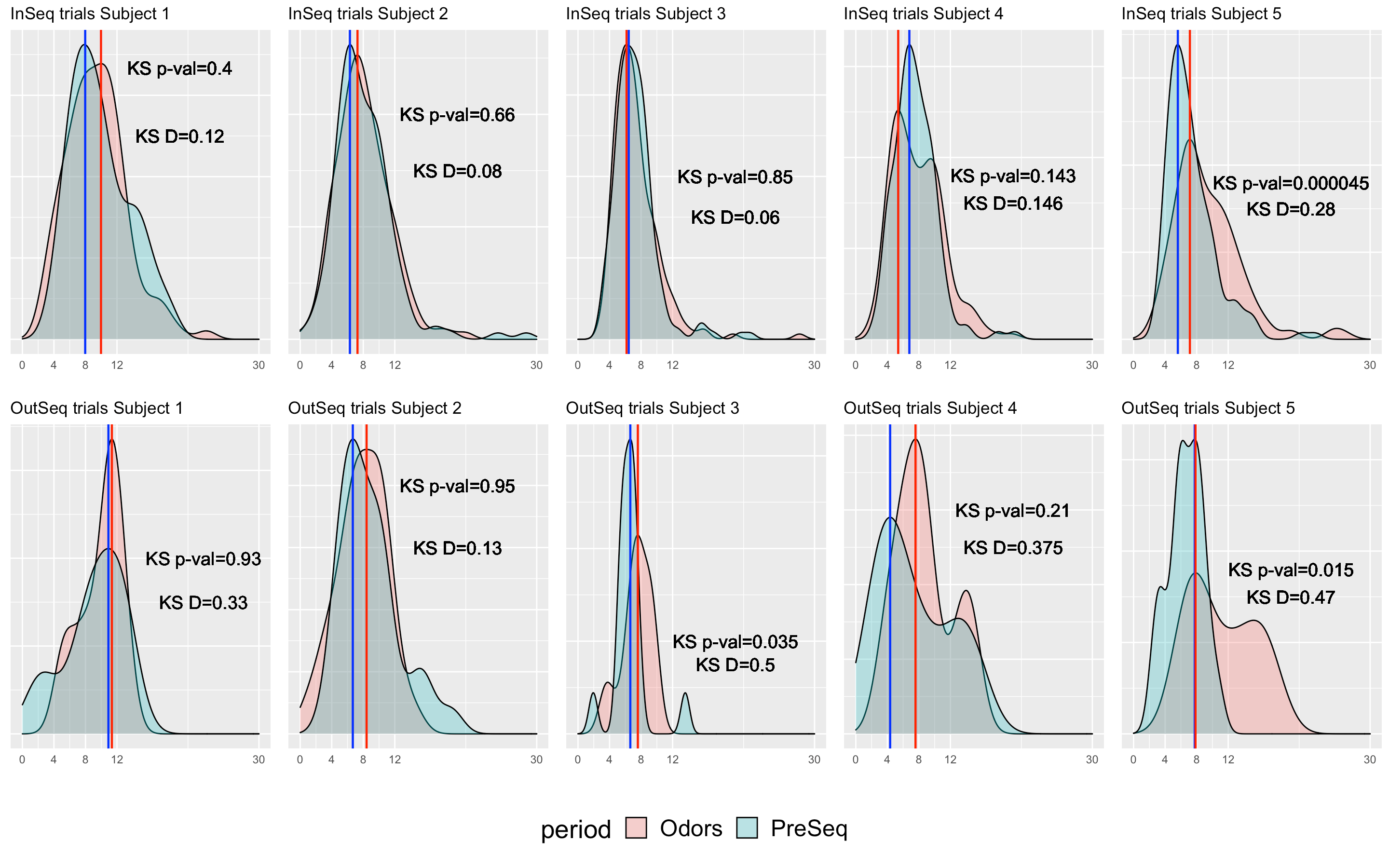}
    \caption{Peak activity distribution for one hour session for all subjects, the vertical lines correspond to the mode of the localized peaks with the highest mixture weight by OutSeq and InSeq contexts accordingly. The right of each graph contains the p-value and test statistic D of the Kolmogorov-Smirnov test.
    }
    \label{5ratsdistr}
\end{figure}

We summarize the distribution of the peak frequency of each subject by considering the AR(2) component with the highest estimated contribution to the variance for each trial. More specifically, for each subject, trial,  
and temporal context, we selected the component with the biggest estimated weight. Usually their values were at least $65\%$ with high concentration around $90\%$. The main-peak activity subject-specific distributions (derived across all trials for each combination of pre vs post-odor and Inseq vs Outseq conditions) are shown in Figure \ref{5ratsdistr}.
We first consider the Inseq trials (top row) for each of the 5 subjects. 
For subject 1 (S1), the distribution of the peak frequency pre-odor is unimodal with support over 0-25 Hertz; for post-odor the distribution of the peak frequencies is also unimodal with the same support. The main difference between the pre-odor and post-odor for the distribution of the peak frequencies during the Inseq condition is the mode: it is 8 Hertz for pre-odor while it is higher at 10 Hertz for post-odor. For subject 2 (S2), the distributions are unimodal for both pre- and post-odor but the support is narrower with concentration on 0-18 Hertz. The mode for the peak frequency for pre-odor is 6 Hertz, while it is 8 Hertz for post-odor. For subject 3 
(S3), the support is even narrower with concentration on 0-12 Hertz and the modes are almost identical for pre-odor and post-odor peak activity at approximately 6 Hertz. 
Subject 4 (S4) has a similar support as subject 1 (S1) but displays a unique feature because the mode for the pre-odor is 7 Hertz which is higher than that for the post-odor which is at 5 Hertz. For subject 5 (S5), the mode for the peak frequency at pre-odor is 5 Hertz vs 7 Hertz for the post-odor. Note that for the Inseq condition we see quite a variation in the brain functional response across the 5 subjects. Indeed for this reason we cannot find a solid justification for developing a single unifying model for these 5 subjects. Thus, we shall proceed with an individual modeling for each subject and describe similarities and differences in the results across the subjects.

We conducted a formal test for the hypothesis of equality of the distributions of pre-odor vs  post-odor using the Kolmogorov-Smirnov method. These tests were conducted separately for each of the 5 subjects and the findings shown on each of the distribution graph in Figure \ref{5ratsdistr}. Under a $5\%$ of significance level, S5 during InSeq trials is the only subject with significant difference of the pre and post distributions (pvalue=$4.5 \times 10^{-5}$). Note that subjects S1, S2, and S3 clearly suggest equality of distributions while for S4 is not significant despite the bimodality of the post-sequence distribution.
Note that the identification of these precise peaks were made possible because the BMARD method gives a representation of the SDF in terms of the building blocks which are the AR(2) spectra. These precise differences in the modes would not have been detected using standard approaches where the frequency bands were predetermined rather than adapted to the specific data that is being analyzed. 
The distribution of peaks for the Outseq condition looks different from the Inseq condition uniformly across the 5 subjects. 
We observe some evidence of bimodality and also a greater visual separation between the distribution of the pre-odor and post-odor presentation, again uniformly across the 5 subjects. Most notable again is S5 (fondly called "Super-Chris" in the laboratory) whose distribution of pre-odor and post-odor peaks have a different support - despite the estimated modes being very similar. For Super-Chris, the distribution of the peaks for pre-odor is very tight from roughly 4-10 Hertz while the post-odor peak has more variation with the spread from approximately 4 - 16 Hertz. As in the Inseq condition,  a Kolmogorov-Smirnov test for the equality of the distributions of peaks for pre-odor vs  post-odor was used. These tests were conducted separately for each of the 5 subjects which results can be found on the side of each graph of the bottom row of Figure \ref{5ratsdistr} where now subjects S3 and S5 display significant differences(S3: pvalue=$0.035$, S5: pvalue=$0.015$) among distributions while subject S1, S2, and S4 show similar test outcomes as in the InSeq trials. 

\begin{figure}[H]
\centering
    \includegraphics[width=14cm]{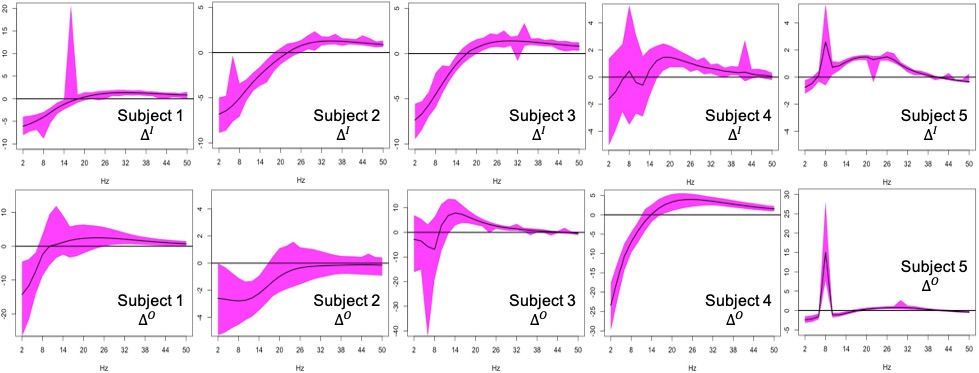}
    \caption{(top row) Functional Boxplot with $95\%$ of internal region computed from 100000 iterations of a resampling scheme on the difference $\Delta^I(\omega)= f^I_A(\omega)-f^I_B(\omega)$ for each subject. (bottom row) Functional Boxplot with $95\%$ of internal region computed from 100000 iterations of a resampling scheme on the difference $\Delta^O(\omega)= f^O_A(\omega)-f^O_B(\omega)$ for each subject.}
    \label{deltascurves5rats}
\end{figure}

To address questions (a.) and (b.) above, we first define the frequency-specific difference in the SDF within InSeq trials and within Outseq trials to be, respectively,
\begin{eqnarray*}
\Delta^I(\omega) & = & f^I_A(\omega)-f^I_B(\omega) \\
\Delta^O(\omega) & = & f^O_A(\omega)-f^O_B(\omega)
\end{eqnarray*}
for all $\omega \in (0,.5)$ where $f^I_A(\omega)$ and $f^I_B(\omega)$
are the SDF for Inseq trials for, respectively, the pre-odor and 
post-odor presentation. Thus,
the quantity $\Delta^I(\omega)$ measures 
the extent of the frequency-specific 
change after the subject detects an odor presented under the setting of correct sequential order. The SDFs for Outseq 
trials the frequency-specific change 
$\Delta^O(\omega)$ 
are defined in a similar manner. The functional boxplots for the differences for the pre-odor vs post-odor for the Inseq and Outseq conditions ($\Delta^I(\omega)$ and $\Delta^O(\omega)$) are given in Figure \ref{deltascurves5rats} where one observes positive differences for higher frequencies, consistent with Figure \ref{5ratsdistr} on the activation of this frequencies after the stimulus is presented to each subject. It is stronger in the InSeq trials shown on the top row. Besides, some peaks on the boxplots between 8 and 14 Hz aligned with the mode of the main peaks locations.  

We now address the question of (c.) interaction between conditions and temporal context. Indeed, a natural question posed by the neuroscientists is whether the change (pre vs post-odor) differs between the Inseq and Outseq conditions. In particular, the task is to identify the specific frequencies (or bands) where the changes (pre vs post-odor) are more highlighted for Outseq trials (and which are 
more emphasized for Inseq trials). To conduct a formal inference on the interaction, we first define
\[
\Delta^{I-O}(\omega) = \Delta^I(\omega) - \Delta^O(\omega).
\]
For a particular frequency $\omega^*$, 
when $\Delta^{I-O}(\omega^*) >0$ then the change in pre vs post-odor for the Inseq condition is greater than that for the Outseq condition. This will be important for identifying physiological features in signals that differentiate between the two experimental conditions. 
In our implementation, the last MCMC 5000 posterior samples were extracted for each of the conditions before described for all trials. 


The use of the proposed "difference of the change" $\Delta^{I-O}(\omega)$ can also be 
interpreted from another point of view. 
For example, when $\Delta^{I-O}(\omega)>0$ then $\Delta^I_i(\omega)>\Delta^O_j(\omega) \Longrightarrow f^I_A(\omega)-f^I_B(\omega)>f^O_A(\omega)-f^O_B(\omega)$. This can be rewritten in another form and thus leads to the interpretation 
\[
f^I_A(\omega)-f^O_A(\omega)>f^I_B(\omega)-f^O_B(\omega).
\]
The quantity $f^I_A(\omega)-f^O_A(\omega)$ 
measures the difference between the spectral power for Inseq vs Outseq conditions during the pre-odor presentation; whereas the $f^I_B(\omega)-f^O_B(\omega)$ measure the difference between Inseq and Outseq during the post-odor presentations.

\begin{figure}[H]
\centering
    \includegraphics[width=14cm]{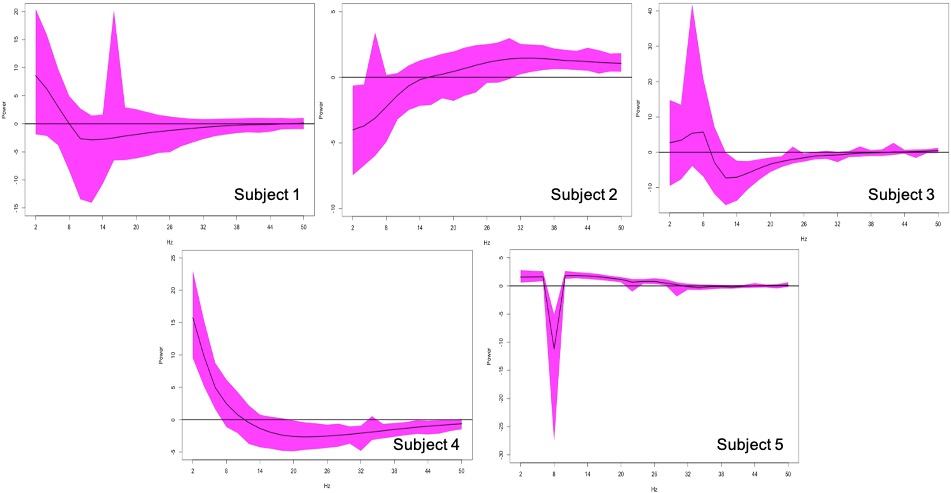}
    \caption{Functional Boxplots with $95\%$ of internal region for the curves $\Delta^{I-O}(\omega)=\Delta^I(\omega)-\Delta^O(\omega)=f^I_A(\omega)-f^I_B(\omega)-(f^O_A(\omega)-f^O_B(\omega))$ showing the median curve. The curves used in the boxplot were computed using a resample scheme of $10^5$ samples from the BMARD posterior SDF curves.
    }
    \label{deltadelta}
\end{figure}

The posterior inference of the curve $\Delta^{I-O}(\omega)$, shown in Figure ~\ref{deltadelta}, displays the functional boxplot from the \textit{fda} package in R set to show the $95\%$ internal region. For subjects S1, S3 and S4, we observe a similar pattern of decay for frequencies higher to 8 Hz but only S3 shows significant differences below 0 for frequencies between 12 and 24 Hz. subject 4 has a significant $\Delta^{I-0}(\omega)$ for $ 22<=\omega<=32$Hz while the significant differences for S4 appear only for $\omega<8Hz$. A different behavior is noted for S2 who shows positive $\Delta^{I-0}(\omega)$ for $\omega>32$Hz.  
Super-Chris (S5) shows quite a distinct pattern. Most notably, the change for pre-vs-post during Outseq is significantly greater than the change during Inseq at a very narrow band around 6-10 Hertz. Thus, the BMARD method produced a highly specific band which identifies the difference in the brain reaction to a correct sequence vs incorrect sequence of the odors.

The potential impact to neuroscience brought by the new findings obtained from BMARD includes are as follows: (a.) identifying the specific frequency (or narrow bands) of the most dominant neuronal oscillations that are 
engaged in memory; (b.) leading to new sets of hypothesis about memory and designing new experiments that test for intervention effects such as applying electrical stimulation at the identified frequency peaks.


\section{Conclusion}


The primary contribution of the BMARD method is that extracts information from the data to provide highly specific frequency information about spectral density function. It gives the estimated number of spectral components (peaks) and provides precise identification of the dominant frequencies where the SDF attains localized peaks. It also gives the corresponding spread (bandwidth) for each of the spectral peaks that were identified. The  determination of the number of components, the  location of the peaks, and the spread are all data-adaptive rather than imposed \emph{a priori} using the standard methods. The Bayesian framework 
facilitates inference on many subject-matter  
hypotheses. 
The construction of the 
SDF under the BMARD method uses a family of  autoregressive kernels which, along with a Dirichlet process prior, gives rise to a Bayesian nonparametric discrete mixture model. BMARD decomposes a stationary univariate process as a linear mixture of latent AR(2) processes, where each component is associated to a unique peak on the SDF. 
The weights of the mixture provide an insight into the components contribution of each latent process to the total variance of the observed signal. Moreover, as demonstrated in the simulations, BMARD gives very good estimates without requiring a higher number of components, which is essential when the sampling rate is low. Thus, even with a relatively few components, the location of the frequencies corresponding to spectral peaks are well estimated because the method data-adaptively identifies the optimal placement of these peak frequencies.

The comparison of BMARD with other approaches points to a limitation when estimating a moving average SDF due to its concavity and smooth shape. Since the AR(2) kernel is a convex function, fitting BMARD to a smooth SDF results in a mixture of several kernels localized by peaks arising from the variability of the periodogram. When the SDF is shaped by sharp peaks as the autoregressive models, BMARD provides parsimonious estimator for the SDF, and outperform other methods. However, despite this limitation, the BMARD still performs very well using the metric of identifying the frequencies that produce the spectral peaks.

The LFP analysis of 5 rats shows how the signal is decomposed into different frequency components during performance on an odor sequence memory task, and how the distribution of peak activity varies across trial types. The significant difference between InSeq and OutSeq trials (in which odors were presented in the correct or incorrect order, respectively) provides compelling evidence that hippocampal LFP activity carries significant information about the sequential organization of our experiences, specifically whether or not events occurred in the expected order. This is an important finding because LFP activity reflects the summed influence of large groups of neurons near the electrode tip. 
In fact, hippocampal oscillations are generally viewed as playing an important role in synchronizing neural activity across neuronal ensembles and circuits, or promoting distinct information processing states (reviewed in \cite{colgin2016rhythms}). Beyond our specific findings, the development of this model may have broader implications in neuroscience as a novel approach to extract additional trial-specific information from LFP recordings, an electrophysiological approach extensively used in the field. 

The decomposition representation in the BMARD method provides a different insight for weakly stationary processes as composed of latent processes 
with various oscillatory behavior. The application of BMARD is broad and could extend well beyond neuroscience. 
It is applicable to other types of data such as weather evolution 
composed of natural cycles with different periodicity or financial data that exhibit economical cycles that are not necessarily sinusoidal but 
can be better represented by simpler stochastic processes explaining short, medium, and long term tendencies.  

\section*{Acknowledgments}
The authors thank Dr. Hart (see \cite{hart2018multi}) for generously sharing his computer codes.
We acknowledge financial support from the KAUST Research Fund and the 
NIH 1R01EB028753-01 to B. Shahbaba and N. Fortin.

\bibliographystyle{chicago}
\bibliography{biblio}

\newpage 

\section*{Appendix: MCMC Algorithm}
We implemented a Metropolis-Hastings within Gibbs to sample from the posterior distribution of the parameters. Our algorithm first updates the number of components with a birth-death process, which at each iteration proposes with equal probability to increase the number of components by one or decrease it by one. In the case of a birth step, the M-H ratio is.
\begin{equation}
q(\theta|\theta^* )/q(\theta^*|\theta )=\big( I(\epsilon_j^{*(i) }<\psi_j^{(i) }  ) (\epsilon_j^{*(i) }-\epsilon_{j-1})^{-1} + I(\epsilon_j^{*(i) }>\psi_j^{(i) }  ) (\epsilon_{j}-\epsilon_j^{*(i) })^{-1} \big)^{-1}  
\label{birth}
\end{equation}
Where $\epsilon^{*(i)}$ represent a new generation for the partition in the interval $( \epsilon_{j-1},\epsilon_{j})$ assuming the random selection of the index $j$ to split that subinterval and $I$ is the indicator function. While in the the case of choosing a death step the M-H probability is computed as 
\begin{equation}
q(\theta|\theta^* )/q(\theta^*|\theta )= I(\epsilon_{j}>\psi_j^{*(i) }  ) (\epsilon_{j}-\epsilon_{j-1})^{-1} + I(\epsilon_{j}<\psi_j^{*(i) }  ) (\epsilon_{j+1}-\epsilon_{j})^{-1}   
\label{death}
\end{equation}
 where $\psi^*$ represents the uniform draw when two component are joined by deleting the value $\epsilon_j$ from the partition. The proposal distribution to update of the location parameters is uniform in the interval $(\psi_c -\epsilon,\psi_c +\epsilon )$  with appropriate conditions to take the modulus over the subinterval defined by the partition.  For the scale parameter, we use a similar uniform draw over the interval $(L_c -\epsilon,L_c +\epsilon)$.
 
 The alpha parameter is updated based on the slice sampling to generate a random walk over the subgraph of the marginal posterior distribution of $\alpha$ given by: 
\begin{equation}
\pi(\alpha | C ) \propto \pi(\alpha) \alpha^{C-1} (\alpha+T) \beta(\alpha+1,T)
\label{alphapos}
\end{equation}
where $\pi(\alpha)$ is the prior over $\alpha$ in our algorithm we set $\pi(\alpha)$ as log-normal, $T$ the observed process size, and $\beta(.)$ is the beta function. The next algorithm presents the steps described.

\begin{algorithm}[H]
\tiny 
\SetAlgoLined
 Propose:$C^{(0)}, M, d$ \;
 Initialize randomly:$\psi_1^{(0)},\dots,\psi_C^{(0)}, L_1^{(0)},\dots,L_C^{(0)},  V_1^{(0)},\dots,V_M^{(0)},\epsilon_1^{(0)},\dots,\epsilon_{C-1}^{(0)}$ \;
 \While{$i \leq MCMC\ chain\ size$}{
  Choose Death or Birth with probability equal to 1/2\;
  \eIf{Death}{
   Choose $j$ randomly with probability $1/C^{(i)}$\;
   Remove $\epsilon_{j}$\;
   Propose a new $\psi_j^{*(i)}$ uniformly on $(\epsilon_{j-1},\epsilon_{j+1})$\;
   Propose a new $L_j^{*(i)}$ uniformly on $(0,d)$\;
   Compute the M-H probability \;
   Make a Reject-Acceptance M-H step for the new $C^{*(i)}=C^{(i-1)}-1, \bar{psi}^*,\bar{L}^*  $ \;
   }{
    Birth: Generate a new component \;
   Choose a $j$ randomly with probability $1/C^{(i)}$\;
   Propose a new $\epsilon_j^{*(i)}$ uniformly on $(\epsilon_{j-1},\epsilon_{j})$\;
   Propose a new $\psi_j^{*(i)}$ uniformly on the interval where is not yet a $\psi$ \;
   Propose a new $L_j^{*(i)}$ uniformly on $(0,d)$\;
   Compute the M-H probability \;
   Make a Reject-Acceptance M-H step for the new $C^{*(i)}=C^{(i-1)}+1, \bar{psi}^*,\bar{L}^*  $ \;
  }
  Sample jointly the vector $\bar{\psi}^{(i)}$\;
  Sample jointly the vector $\bar{L}^{(i)}$\;
  Sample jointly the vector $\bar{V}^{(i)}$\;
  Sample jointly the vector $\bar{Z}^{(i)}$\;
  Sample $\alpha^{(i)}$\;
 }
 \caption{M-H within Gibbs DP AR(2) mixture for stationary processes}
\end{algorithm}




\end{document}